\documentclass[letterpaper]{jpconf}
\usepackage{graphicx}

\begin{document}
\nocite{*}

\title{First CMS Results with LHC Beam}

\author{Toyoko J. Orimoto, on behalf of the CMS Collaboration}

\address{California Institute of Technology, 1200 E. California Blvd, Pasadena, CA 91125, USA}
\ead{toyoko@hep.caltech.edu}

\begin{abstract}
The Compact Muon Solenoid (CMS) Experiment is a general purpose particle detector experiment located at the Large Hadron Collider (LHC) at CERN.  In 2008, the LHC beam was commissioned and successfully steered through the CMS detector.  First results from CMS with beam data are described, focusing on detector commissioning with beam data, from beam halo studies with the endcap muon system to displays of ``beam splash'' events, in which the proton beam was stopped by an upstream collimator.
\end{abstract}

\section{Introduction}

\begin{figure}[bt!]
\begin{center}
\includegraphics[height=5.5cm]{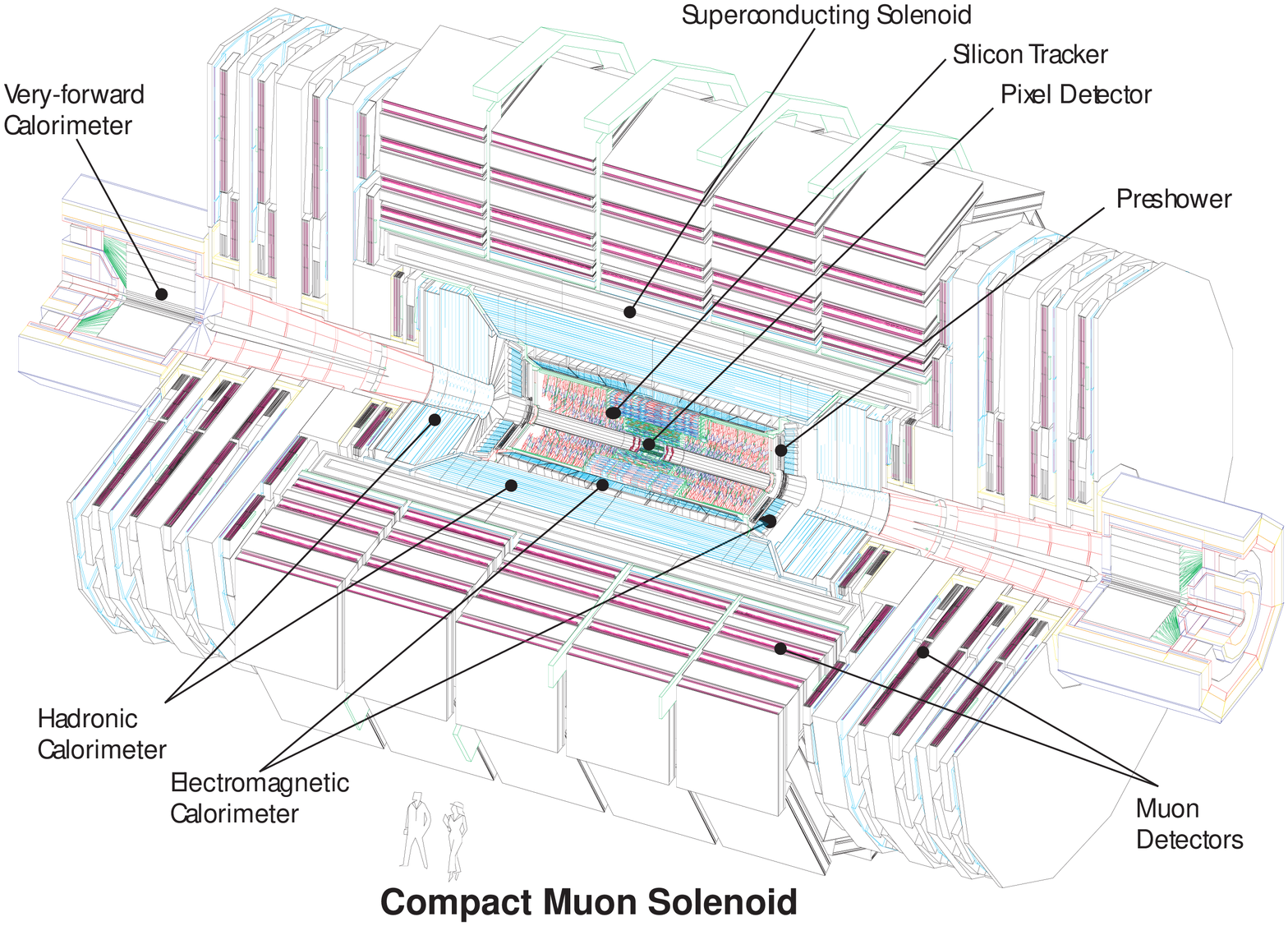}\hspace{1pc}
\includegraphics[height=5.0cm]{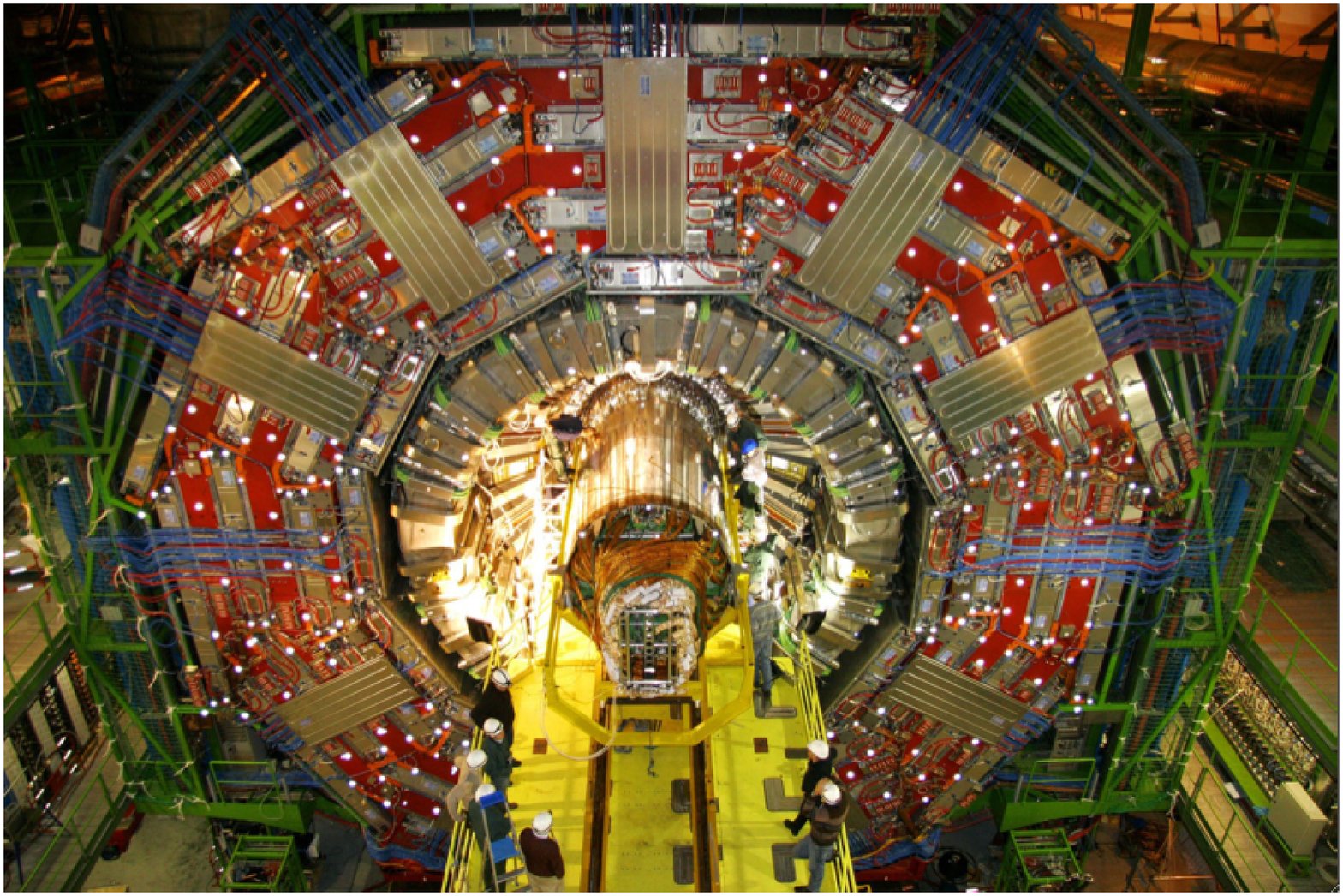}
\caption{\label{fig:cms}(left) Schematic of the Compact Muon Solenoid detector. (right) The CMS detector during installation of the silicon strip tracker, Dec 2007.  }
\end{center}
\end{figure}

The Compact Muon Solenoid (CMS) Experiment is a general purpose particle detector experiment located at the Large Hadron Collider (LHC) at CERN, in Geneva, Switzerland.
The design of the  CMS detector centers around a 3.8T iron-core superconducting solenoid.  Within the solenoid, at the very heart of the detector, is an all silicon tracker, consisting of 66M pixel and 10M strip detectors.  
Calorimetry is performed by a high-resolution, high-granularity electromagnetic calorimeter (ECAL), comprised of more than 70k lead-tungstate crystals, and a brass-scintillator hadronic calorimeter (HCAL).
Just outside the solenoid is a scintillator-based tail-catching hadronic outer calorimeter (HO).  In addition, a steel and quartz fibre forward hadronic calorimeter system (HF) is placed on both sides of the detector endcaps to ensure hermiticity.
The iron yoke of the magnet system is instrumented with a muon spectrometer, composed of 3 detector technologies: drift tubes (DT) in the central region and cathode strip chambers (CSC) in the endcaps, both complemented by resistive plate chambers (RPC).
Considering the late availability of the underground cavern, a modular detector design was chosen to allow for construction, installation, and testing before lowering underground.  
%
Figure~\ref{fig:cms} (left) shows a schematic drawing of the CMS Detector.
Further detail on the design and performance of the CMS detector can be found elsewhere~\cite{cmsjinst,tdr}.

The construction phase of the CMS detector is now complete, and the detector has been fully installed and commissioned in the underground cavern at LHC Point 5.   
Figure~\ref{fig:cms} (right) shows a photo of the CMS detector during installation of the tracker.
Since May 2007, CMS has undergone frequent ``global runs'' devoted to global commissioning, exercising all of the installed detector subsystems together.
In addition to testing the global readout, global runs have also helped exercise event selection and reconstruction, data analysis, alignment and calibration, data quality monitoring, computing operations and data transfer.
The complexity and scope has increased with each global run, with a larger fraction of subdetectors joining at each interval.
The global runs culminated in data taking for over a week with first LHC beam in September 2008, and a two week Cosmic Run at Full Tesla (CRAFT) in October-November 2008.

%
%
%

\section{Performance with First LHC Beam Data}\label{sec:beam}

In September 2008, the CMS detector collected first beam data from the LHC.  The first proton beams were circulated successfully on 10 September 2008.  Single beams were circulated through the LHC for several days, at injection energy of 450 GeV, until 19 September.  During this time, CMS triggered and recorded data successfully.

\begin{figure}[bt!]
\begin{center}
\includegraphics[height=4.5cm]{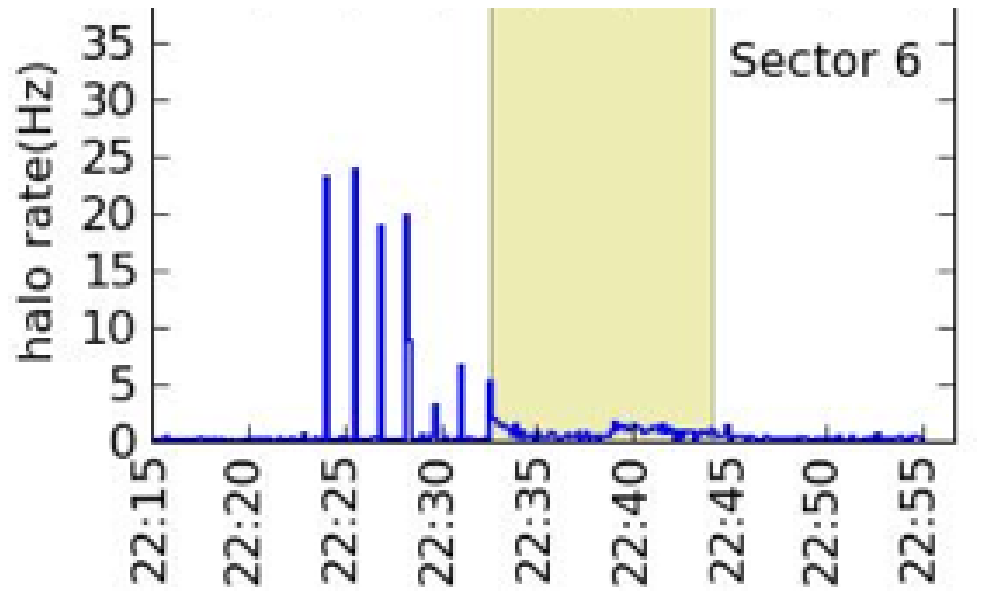}
\includegraphics[height=5.0cm]{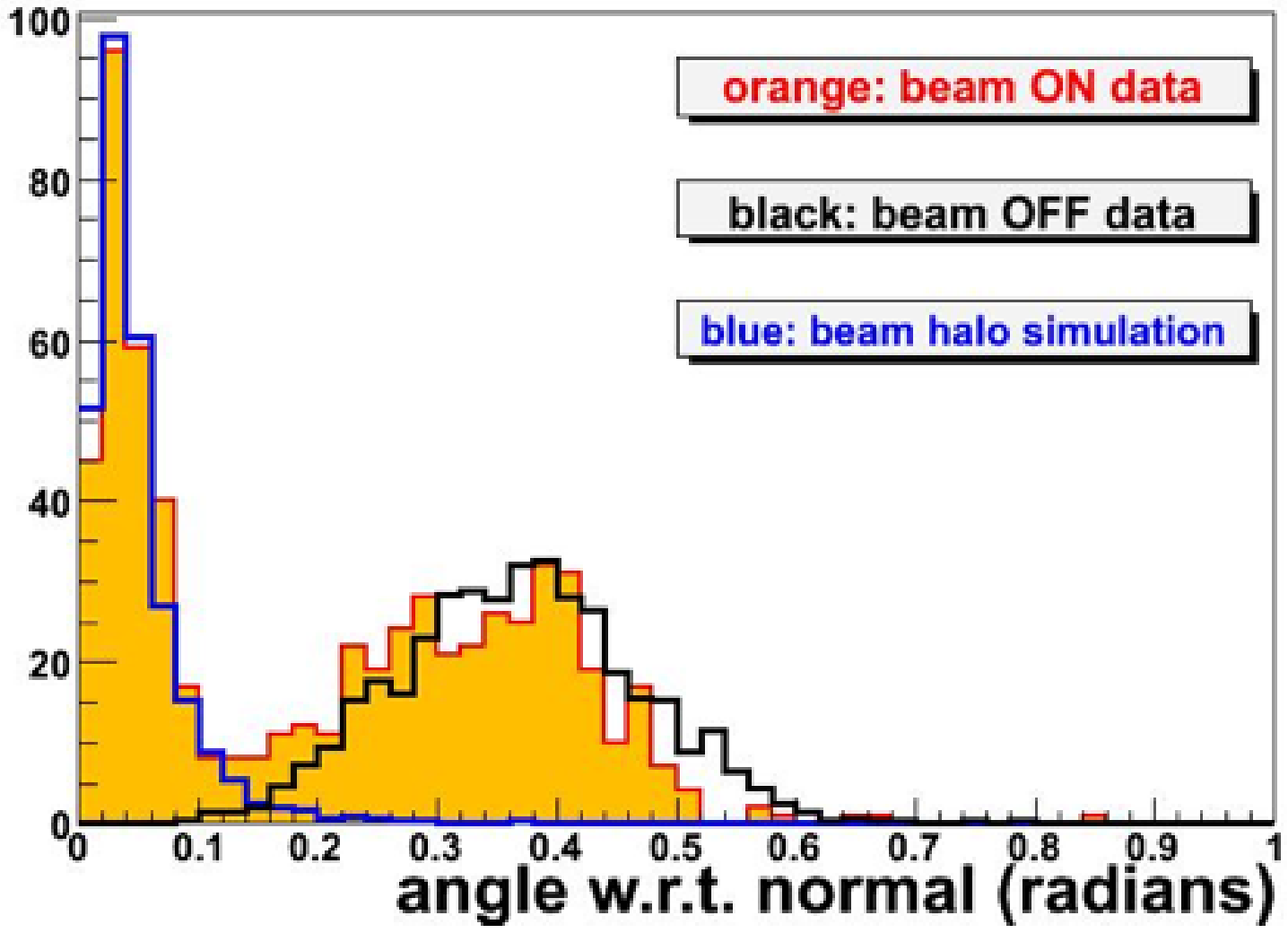}
\caption{\label{fig:csc}(left) Beam halo rate (Hz) as a function of time (hour:min).  The shaded region denotes the ~10min period when the LHC beam was first RF captured.  (right) Angle of muon tracks with respect to beam line, as measured by the endcap muon cathode strip chambers, with beam data (orange filled), cosmics data (black), and beam halo simulation (blue). }
\end{center}
\end{figure}

During beam running, CMS triggered on beam halo muons, muons outside of the beam-pipe arising from decays of pions created when off-axis protons scrape collimators or other beam-line elements.  
Beam halo events were particularly useful for testing the triggering and reconstruction of events in the endcap muon system.  
Figure~\ref{fig:csc} (left) shows the beam halo trigger rate as measured by one sector of the muon endcap CSCs.  The shaded region denotes the period of first successful LHC beam RF capture, which lasted for approximately 10 minutes and ended with a beam abort.  The halo rate measured by the CSCs was clearly reduced during and after the RF capture of the beam.  The HCAL endcap system also saw a ``cleaner'' beam with less energy deposition along the beam-line after the RF capture of the beam.

Figure~\ref{fig:csc} (right) shows the distribution of the angle with respect to the beam-line of muon tracks passing through the endcap CSC system.
Since beam halo muons pass through the detector collinear with the beam-line, we expect this angle to be small for pure beam halo events (beam halo simulation, in blue).  Cosmic muons pass through the detector more orthogonally to the beam-line, and thus we expect this angle to be larger for cosmic muon ``beam off'' events (in black).  For ``beam on'' events, we expect a combination of beam halo and cosmic muon events (in orange filled).  Figure~\ref{fig:csc} shows this trend validated with data and simulation events.

\begin{figure}[b!]
\begin{center}
%
%
\includegraphics[width=0.49\textwidth]{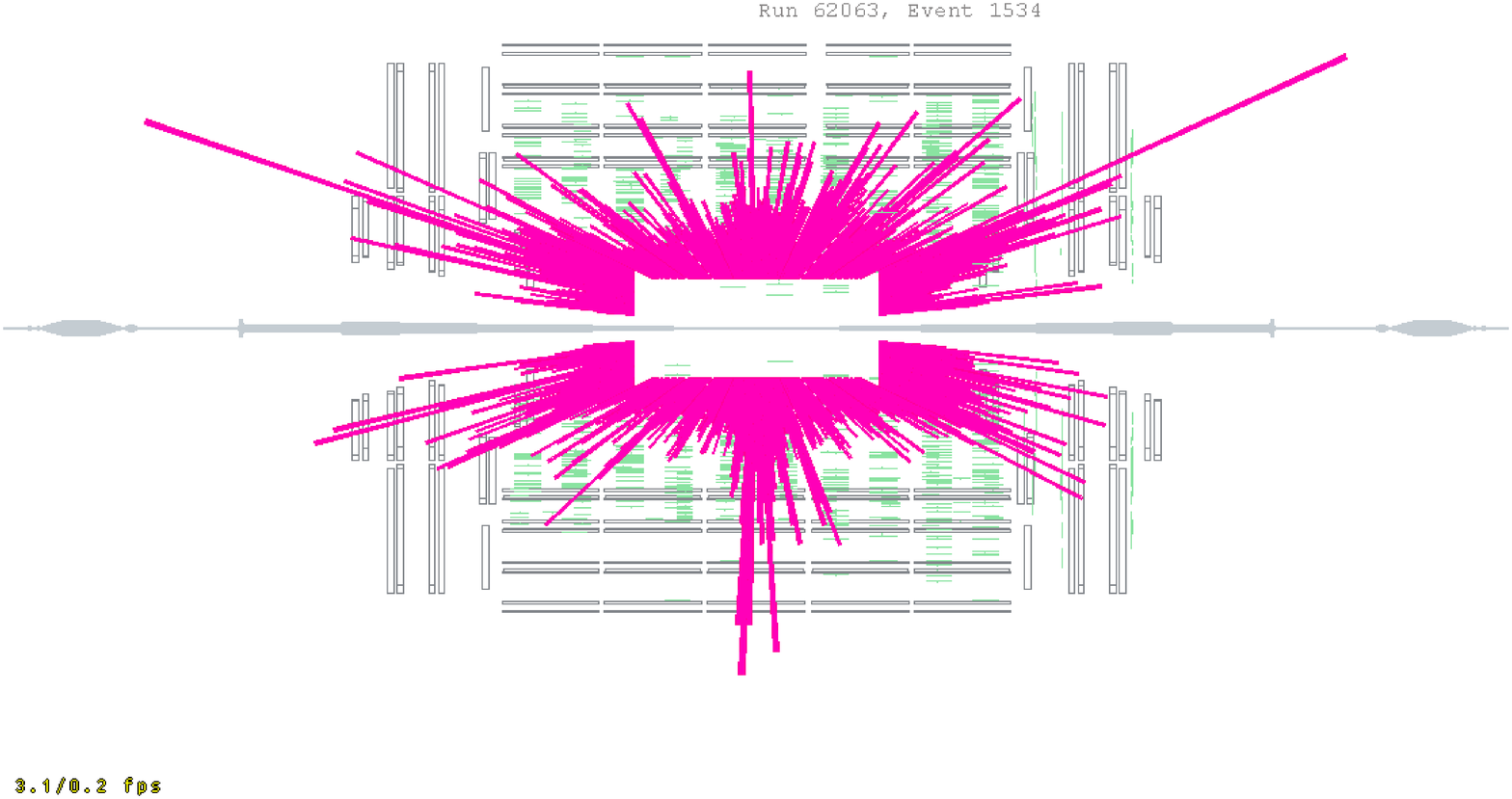}
\includegraphics[width=0.49\textwidth]{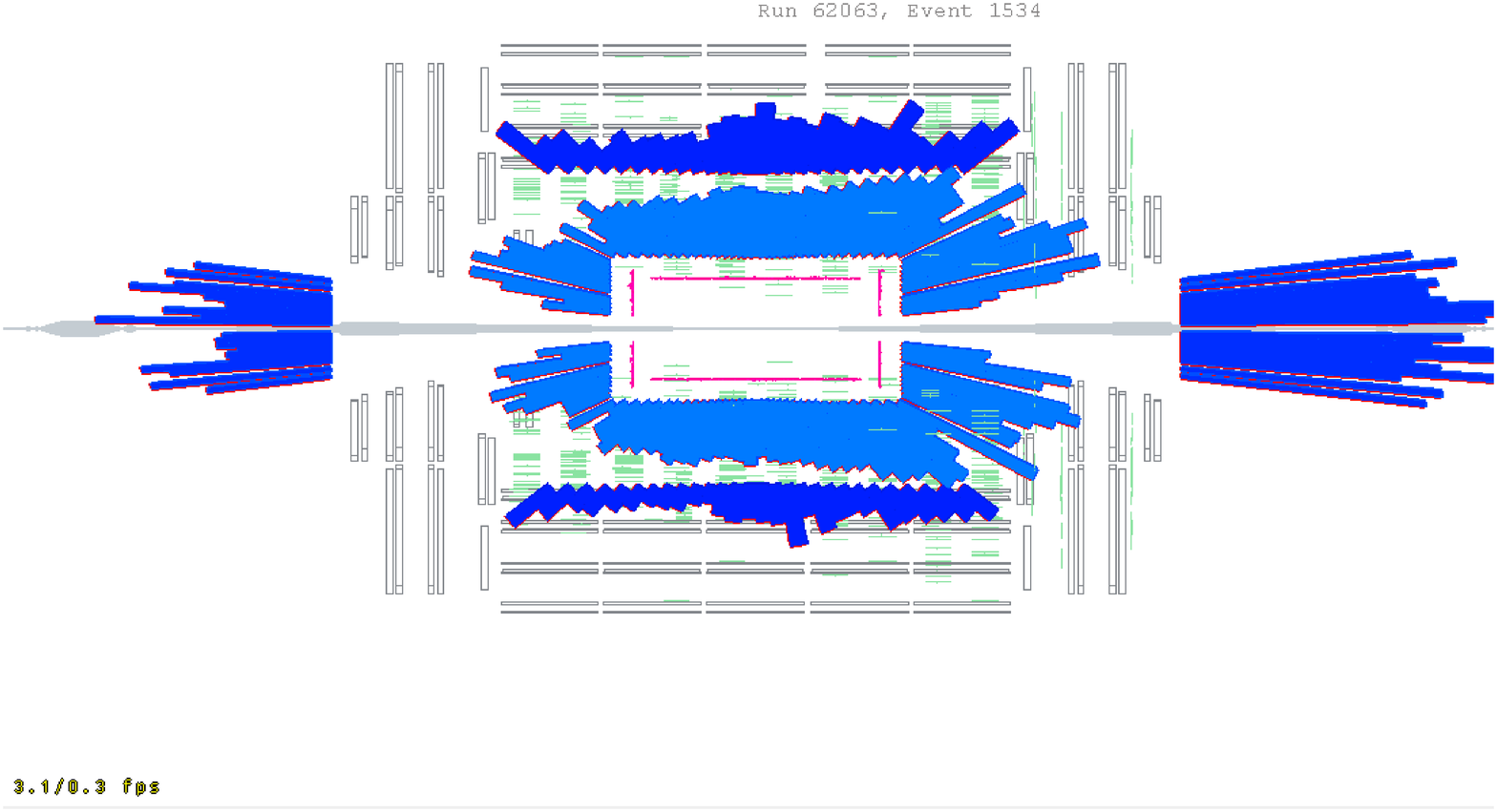}
\caption{Event display images from a single ``beam splash'' event.  
The left image shows the energy deposited in the electromagnetic calorimeter (pink).  The right image shows the energy deposited in the hadronic calorimeter (blue).
In a single beam splash event, $\sim100$ and $\sim1000$ TeV of energy were deposited in the hadronic and electromagnetic calorimeters, respectively. 
\label{fig:evtdisplay}}
\end{center}
\end{figure}

In the days preceding September 10th, single beam shots were sent onto closed collimators near the CMS detector.  The interactions of the beam with the collimator and surrounding material created a large number of secondary particles which passed through the CMS detector.  These beam-on-collimator events are called ``beam splash'' events because of the spectacular signature they left in the detectors.  In particular, these events deposited an enormous amount of energy in the calorimeters;  in a single beam splash event, $\sim100$ and $\sim1000$ TeV of energy were deposited in the hadronic and electromagnetic calorimeters, respectively.  Figure~\ref{fig:evtdisplay} shows an event display for such a ``beam splash'' event. In addition to leaving a spectacular signature in the detector, these events have been used to internally synchronize ECAL and HCAL, and also to validate the inter-calibration of the ECAL endcap channels.
%

%

\section{Performance with Data from the Cosmic Run at Full Tesla (CRAFT)}\label{sec:craft}

The aim of the Cosmic Run at Full Tesla (CRAFT) was to run CMS for four weeks, with all subsystems participating, collecting data continuously to further gain operational experience before data-taking with $p$-$p$ collisions.  In addition, a major goal of CRAFT was to operate CMS at full field (3.8T) for as much of the running period as possible and to study the effects of the magnetic field on the detector components.  Prior to CRAFT, the most extensive test of the CMS detector at full magnetic field was during the Magnet Test and Cosmic Challenge in 2006.

\begin{figure}[bt!]
\begin{center}
\includegraphics[height=14pc]{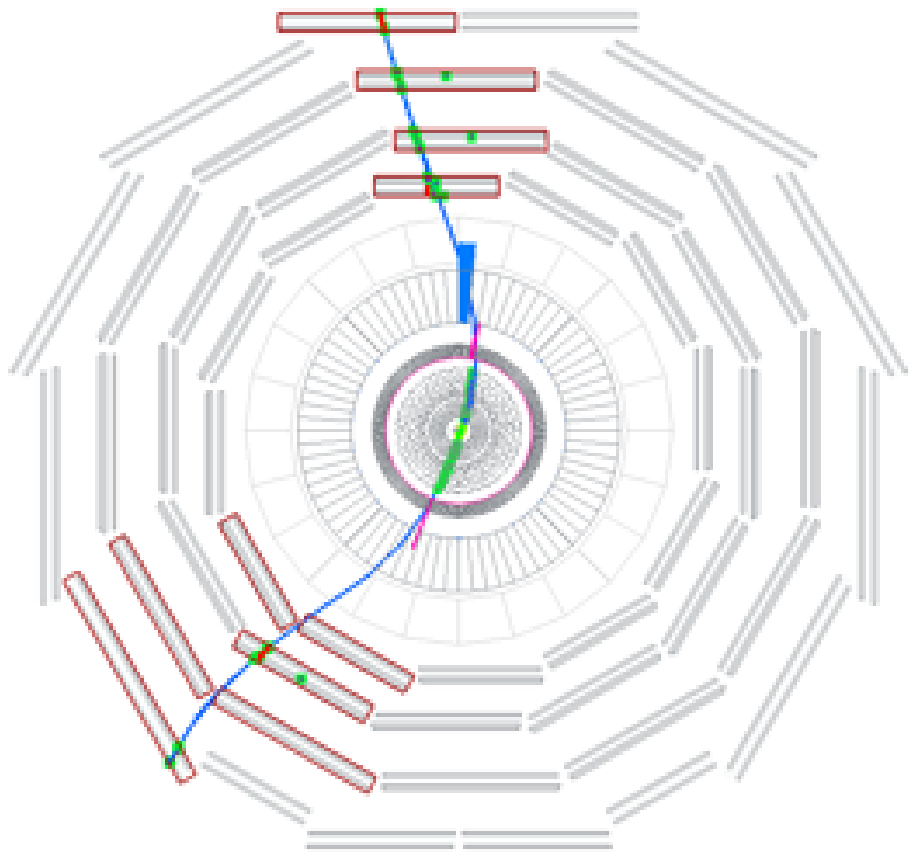}
\includegraphics[height=14pc]{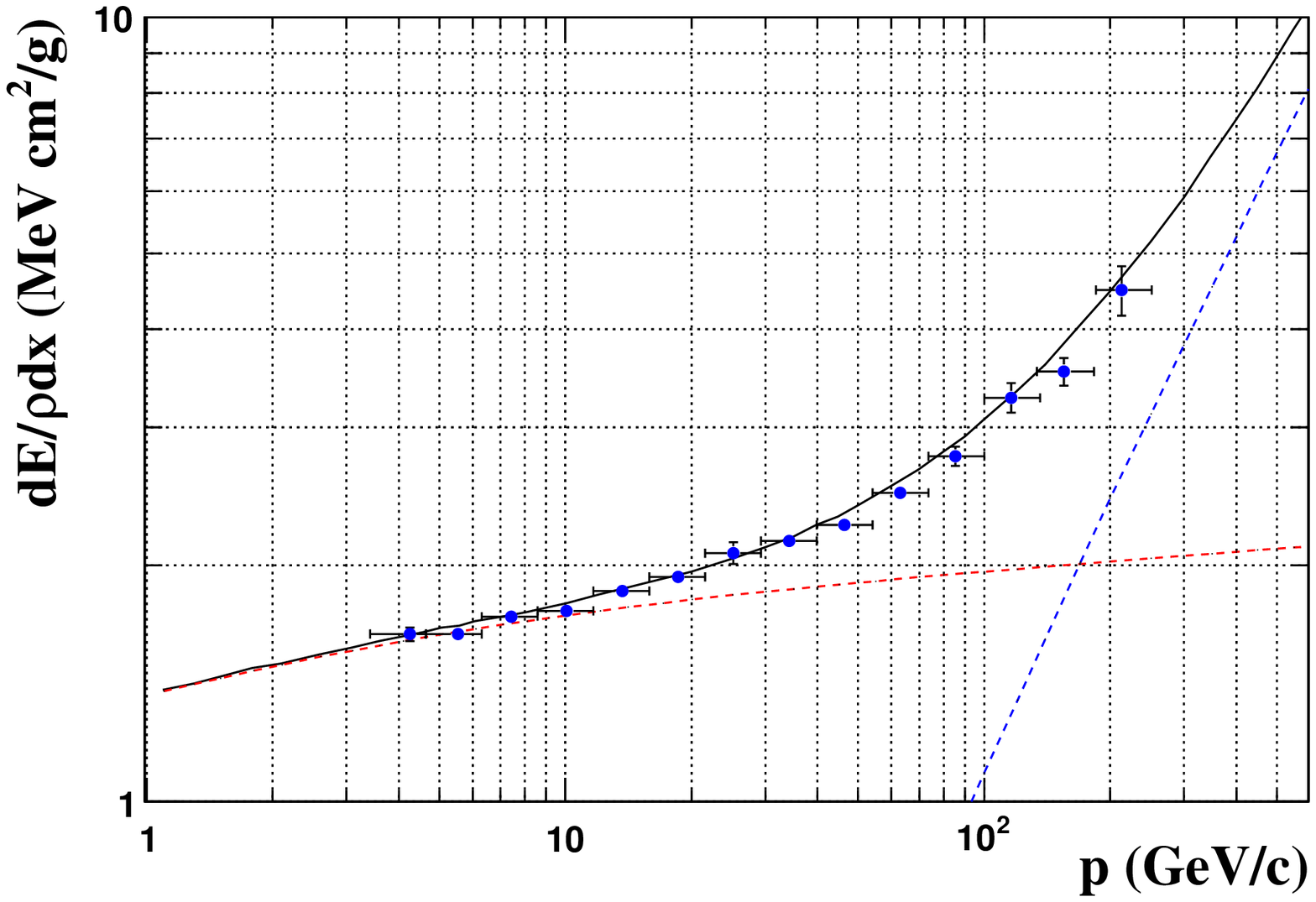}
\caption{\label{fig:craft}(left) Event display for a CRAFT cosmic muon passing through the CMS Detector. In blue is the reconstructed muon track passing through the muon system, the hadronic and electromagnetic calorimeter, as well as the tracker. (right) The $dE/\rho dx$ stopping power of the CMS electromagnetic crystal calorimeter, as a function of muon momentum.  The blue points correspond to data from CRAFT cosmic muon events; the black curve is the expected stopping power of lead-tungstate crystals.  The red dashed line shows the contribution from collision loss; the blue dashed line shows the contribution from bremsstrahlung.}
\end{center}
\end{figure}

During CRAFT, we collected more than 370M cosmic events during four weeks in the interval of 13 October to 11 November 2008.  In this period, CMS was operated with B=3.8T for 19 days.  290M of the 370M events had B=3.8T, with the silicon strip tracker and the muon drift tube system in the readout.  194M of the total had all subdetector components in the readout.  Figure~\ref{fig:craft} (left) shows an event display for a CRAFT cosmic ray muon passing through the CMS detector.
The large CRAFT data set has given rise to many analyses, from subsystem and trigger performance studies, calibration and alignment, validation of the magnetic field map, measurements of cosmic muon flux and charge ratio, to cosmic muon background studies for LHC running.  Figure~\ref{fig:craft} (right) shows the results from one such analysis, the study of energy deposition in the ECAL crystals.  We see good agreement between the measured $dE/\rho dx$ in ECAL and the expected stopped power for lead-tungstate crystals.

%


\section{Conclusions}

After many years of design and construction, the CMS detector has been commissioned and has collected first data with LHC beams.  The detector performance has been proven with beam splash, beam halo, as well as cosmic ray muon events with and without B field.
Since the beginning of September 2008, all installed subsystems of the CMS detector have routinely been in global readout, and the stability of running with all CMS components has been proven.  
Since the CRAFT run at the end of 2008, CMS has continued global commissioning with cosmic ray muons, in preparation for upcoming LHC beams in 2009.

\section*{References}

\end{document}